\documentclass[prl,twocolumn,amsmath,amssymb]{revtex4}

\usepackage{amsbsy}
\usepackage{graphicx}
\usepackage{bm}
\usepackage{graphics}

\renewcommand{\theequation}{\thesection.\arabic{equation}}
\def\spazio#1{\vrule height#1em width0em depth#1em}
\def\enr{E_{{}_{\mathrm{NR}}}}

\begin{document}

\begin{abstract}
{ We study the Dirac equation in  confining potentials with pure vector coupling, proving the existence of metastable states with longer and longer lifetimes as the non-relativistic limit is approached and eventually merging with continuity into the Schr\"odinger bound states. We believe that the existence of these states could be relevant in high energy model construction and in understanding possible resonant scattering effects in systems like Graphene. We present numerical results for the linear and the harmonic cases and we show that the the  density of the states of the continuous spectrum is well described by a sum of Breit-Wigner lines. The width of the line with lowest positive energy, as expected, reproduces very well the Schwinger pair production rate for a linear potential: we thus suggest a different way of obtaining informations on the pair production in unbounded, non uniform electric fields, where very little is known.

}
\end{abstract}

\bigskip
\bigskip

\title{\bf States of the Dirac equation in confining potentials.}

\author{Riccardo Giachetti}
\affiliation{Dipartimento di Fisica, Universit\`a di
Firenze, Italy}
\affiliation{Istituto Nazionale di Fisica Nucleare, Sezione di Firenze}
\email{< name > @ fi.infn.it}
\author{Emanuele Sorace}
\affiliation{Istituto Nazionale di Fisica Nucleare, Sezione di Firenze}

\maketitle

\medskip
\noindent

\bigskip

%
%
%
%


\renewcommand{\theequation}{\arabic{equation}}
\setcounter{equation}{0}


The absence of bound states for the Dirac equation in confining  potentials
poses a delicate question of physical interpretation.
It was in fact shown in \cite{Plesset} that
the asymptotically oscillating behavior of the solutions of 
the Dirac equation with vector coupling 
and a potential given by a positive power of the modulus of the position variable
 implies a purely continuous spectrum. 
The result has been subsequently confirmed  
and  generalized by proving that even any self-adjoint extension of the boundary value problem 
has only a purely continuous spectrum,  \cite{Thaller}.
This situation is contrary to the physical intuition and it makes hard to 
justify a perturbative approach to the relativistic corrections, since no definite perturbed eigenvalues exist.
Most of later investigations have therefore tried different ways to introduce confining potentials into the Dirac 
equation, \textit{e.g.} by scalar coupling or by projection onto the large component
(see \cite{Rui} for a review): doing like that, however,  physically relevant systems as charged particles in strong electric fields
do not find an appropriate description. 
A possible way out of this difficulty was given in the mathematical paper \cite{Tit2}, where
a so called `weak quantization'  was introduced to treat
the  (1+1)-dimensional Dirac equation  with a linear potential.
The analysis is developed in the complex plane of the energy;  when the Schr\"odinger limit is approached it is shown that 
the real part of the energy converges to the  non-relativistic spectrum and the imaginary part becomes exponentially 
vanishing. For a linear potential  analytical solutions are available in terms of special functions and in \cite{Tit2} the spectral quantities were estimated by perturbative expansions, having thus a very limited range of validity, out of which no general picture can emerge. Moreover a coherent physical interpretation was clearly outside the mathematical purpose of the author. To our knowledge no further development along these lines is found in literature since.

From a physical point of view, the presence of an unbounded increasing potential brings to bear
upon the problem arguments similar to those of the Klein paradox, widely studied  
both in first and in second quantization, \cite{Calo,Gre,Cheon}.
Recently a field-theoretical interpretation using numerical methods based on spatial and temporal resolution was given in \cite{Kre}, where it is found that the pair production by the potential is suppressed when the spatial density of the incoming electron overlaps with the potential region and that the transmitted portion of the wave packet, in a single particle description, corresponds  to the amount by which the electron reduces positron's spatial density.
 Although in this letter we  strictly remain in a first quantized framework and we are dealing with the stationary problem defined by the Dirac equation rather than with a scattering picture, still we shall see that  the pair production rate is recovered in a natural way.
Our treatment follows the classical methods of the spectral analysis, \cite{Hille,Tit1},  and, in particular, it deals with  an accurate evaluation of the density of the states of the continuous spectrum. It can be compared with  the phenomenological approach in terms of Gamow vectors, \cite{Gadella}, mainly used to describe resonances in composite systems of solid state and nuclear physics. The investigation is necessarily numerical: thus it does not suffer the limitations of the perturbative expansion and it can easily be extended to more general potentials for which analytical solutions do not exist. We will present in detail the results of the linear and quadratic potentials.
The general situation can be summarized as follows. In non-relativistic quantum mechanics the spectrum is discrete, the  eigenvalues correspond to the real zeroes and poles $\lambda_i$ of a spectral function introduced by Weyl (traditionally denoted by $m(\lambda)$, $\!$\cite{Hille}) and the density of the states reduces to a sum of $\delta$-functions, one for each bound state. In the Dirac equation the $\lambda_i$  move off the real axis, the spectrum becomes purely continuous and the density of the states, $\rho(\lambda)=-\Im( m(\lambda))+\Im(1/m(\lambda))$ for real $\lambda$, appears now as a sum of   Breit-Wigner (BW) lines whose central values, determined by  $\Re(\lambda_i)$, are closer and closer to the non-relativistic eigenvalues and whose widths, determined by  $\Im(\lambda_i)$, are more and more narrow for decreasing values of the ratio of the  interaction to the mass energy. 

The physical interpretation suggested by these facts is that the broadening of the $\delta$-lines is due to transitions between positive and negative energy sectors induced by the supercritical field: thus, although in the relativistic context the nature of the spectrum is completely changed, still narrow BW lines signify the presence of metastable states, giving rise to resonances in the scattering cross section around the line energies. On the one hand, therefore,  the continuity to the  non-relativistic states is  recovered. On the other, contrary to what occurs for the Schr\"odinger ground state, in the relativistic regime also the lowest positive energy state decays. The second quantized counterpart of this fact is that the Fock  vacuum  will not remain such forever, but, according to  the usual theory of the effective action, \cite{Gre,DTZ},  it will decay with the exponential law $|\langle 0(t)|0(t+T)\rangle|^2=\exp(-VT\,w^f)$, where $w^f$ is the pair production rate for unit volume and unit time. We  thus expect similar behaviors of the  line width of the lowest positive energy state and of the pair production \textit{vs.} the interaction strength.  This circumstance appears very well verified for the pair production rate in a constant electric field, as obtained by Schwinger \cite{Schw,Nik}, so that we are led to assume that the width of the first resonance can provide a quantum mechanical way of estimating the pair production for general situations where little is known from QED, as in the case of  unlimited growing potentials (see \cite{Don,Kim} for recent developments).   We then present new data for a quadratic potential, corresponding to an electron in a uniformly growing electric field, finding a pair production behavior very similar to the Schwinger's one.
 We finally believe that our results can be relevant not only in model building of quark systems, \cite{Martin,Crater}, but also in investigations of Klein paradox in strong crystalline fields \cite{Ugg} as well as 
in the very recent and expanding subject of the Graphene physics, where the influence of  impurities is described by the Dirac equation with vector coupling, \cite{Kats}: the  metastable states may result essential for understanding possible effects of resonant scattering.

Consider the (1+1)-dim Dirac equation in a unit system with $\hbar=1$,  
{
\begin{eqnarray}
\psi'(x)-\bigl[\,(1/c)\bigl(E-U(x)\bigr)\,i\sigma_y+mc\,\sigma_x\,\bigr]\,\psi(x)=0
\label{DiracEquation}
\end{eqnarray}
}
\noindent where $\psi(x)={}^t\bigl(\psi_1(x),\psi_2(x)\bigr)$ and $\sigma_i$ are the Pauli matrices.
For the family of even potentials $U(x)=a\,|x|^n$ the equation (\ref{DiracEquation}) can be studied in $[0,\infty)$, having infinity as the unique singularity in the limit point case \cite{Hille}. Therefore, from the Weyl general theory of singular boundary value problems,  \cite{Hille,Tit1}, for $\Im(E)>0$ there exists only one normalizable solution $\widetilde\psi(x)$ of the equation, up to a constant factor. If  $\{\psi^{(i)}(x,E)\}_{i=1,2}$ is a fundamental system  of solutions of (\ref{DiracEquation}) with $\psi^{(i)}_j(0,E)=\delta^i_j$,  the Weyl function $m(E)$ is defined by an expansion $\widetilde\psi(x,E)=\psi^{(1)}(x,E)+m(E)\,\psi^{(2)}(x,E)$. Using the initial conditions, $m(E)$ can be written as
\begin{eqnarray}
m(E)=\widetilde\psi_2(0,E)/\widetilde\psi_1(0,E)
\label{emme}
\end{eqnarray}
Finally the density of the states for real $E_0$ reads, \cite{Tit1},
\begin{eqnarray}
 \rho(E_{0})=\!\!\lim\limits_{\nu\rightarrow 0+} \Bigl( -\Im\bigl(m(E_{0}+i\nu)\bigr)\! +\Im\bigl(1/m(E_{0}+i\nu)\bigr) \! \Bigr)
\label{rhodilambda}
\end{eqnarray}
For calculation reasons, we find it convenient to define
\begin{eqnarray}
\phi(x)={}^t\bigl(\phi_1(x),\phi_2(x)\bigr)=2^{-\frac12}\,i\,(\sigma_y+\sigma_z)\,\psi(x)
\label{defiPhi}
\end{eqnarray}
obtaining for $\phi(x)$ the following equation:
{
\begin{eqnarray}
\phi'(x)-\bigl[\,(i/c)\bigl(E-U(x)\bigr)\,\sigma_z-mc\,\sigma_x\,\bigr]\,\phi(x)=0
\label{equaPhi}
\end{eqnarray}
}
We denote by $\{\phi^{(i)}(x,E)\}_{i=1,2}$  the fundamental  solutions of  (\ref{equaPhi}) with initial conditions $\phi^{(i)}_j(0,E)=\delta^i_j\,$.
Hence, if $\widetilde\phi(x,E)$ is the normalizable solution of (\ref{equaPhi}) corresponding to $\widetilde\psi(x,E)$ and we
expand  $\widetilde\phi(x,E)=\phi^{(1)}(x,E)+\kappa(E)\phi^{(2)}(x,E)$, we then have a finite limit
\begin{eqnarray}
\kappa(E)=-\lim_{x\rightarrow\infty} \phi^{(1)}_i(x,E)/\phi^{(2)}_i(x,E),\quad i=1,2.
\label{kappa}
\end{eqnarray}
The relation between $m(E)$ and $\kappa(E)$ is given by
\begin{eqnarray}
 m(E)=i\,\bigr(\kappa(E)+i\bigl)\,\bigr(\kappa(E)-i\bigl)^{-1}
\label{rel_m_kappa}
\end{eqnarray}
Finally, introducing the ``non-relativistic energy'' $\enr$ by $E=\enr +mc^2$, the elimination of $\phi_2(x)$ yields the second order equation for $\phi_1(x)$
\begin{eqnarray}
 \phi''_1(x)+\bigl[\,2m\bigl(\enr -U(x)\bigr)+c^{-2}R(x) \,\bigr]\,\phi_1(x)=0
\label{equa2Phi1}
\end{eqnarray}
where $R(x)= ic\,U'(x)+(\enr -U(x))^2  $. The non-relativistic limit for $c\rightarrow\infty$  is then evident.

\begin{figure}
\includegraphics*[height=4.8cm,width=7.2cm]{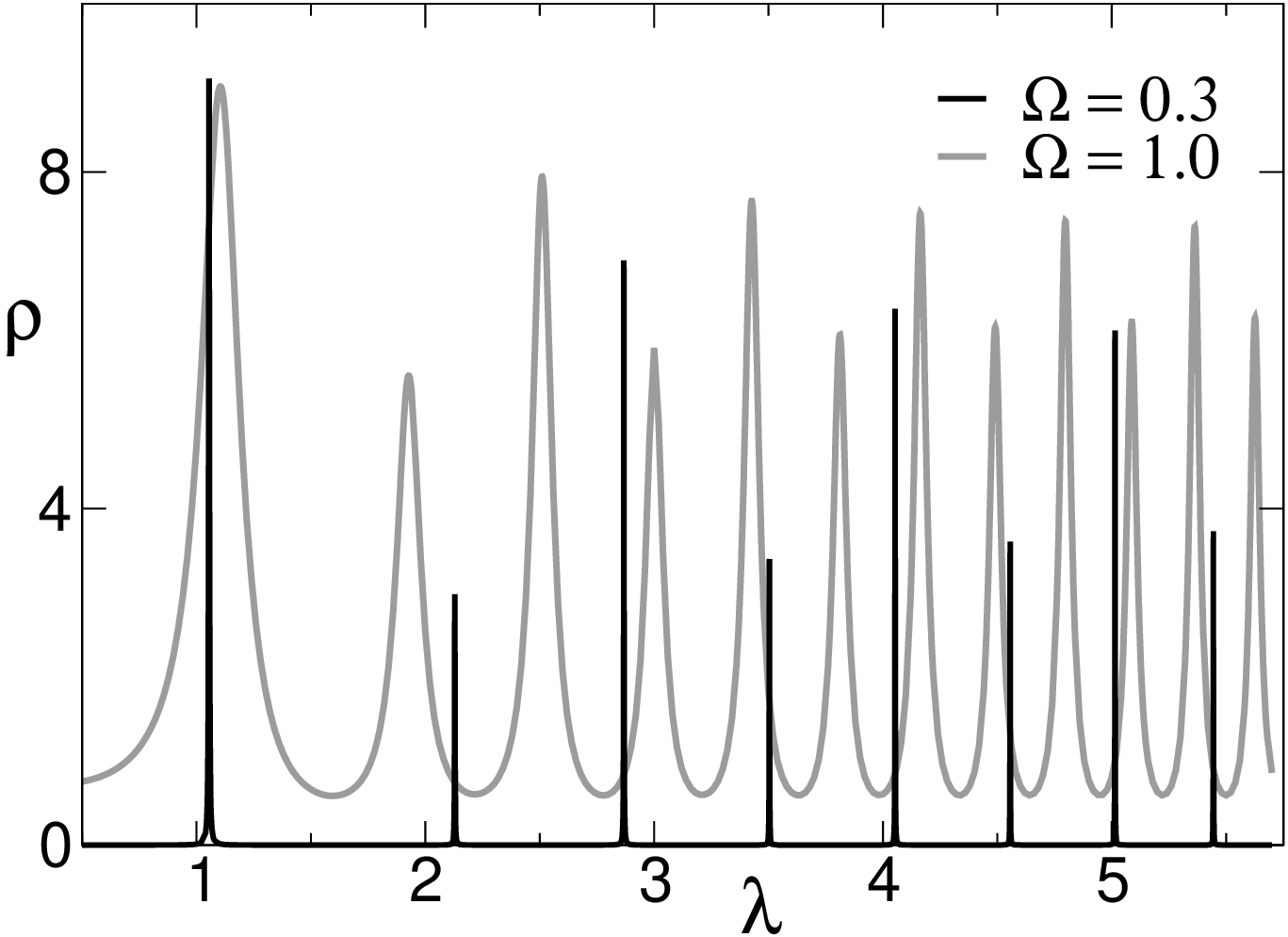}
\caption{The density of the states for the linear potential with $\Omega=0.3,\,1$. The scale
for $\Omega=0.3$ must be multiplied by $10^2$.}
\label{linearspectra}
\end{figure}

\begin{figure}
\includegraphics*[height=4.8 cm,width=8.cm]{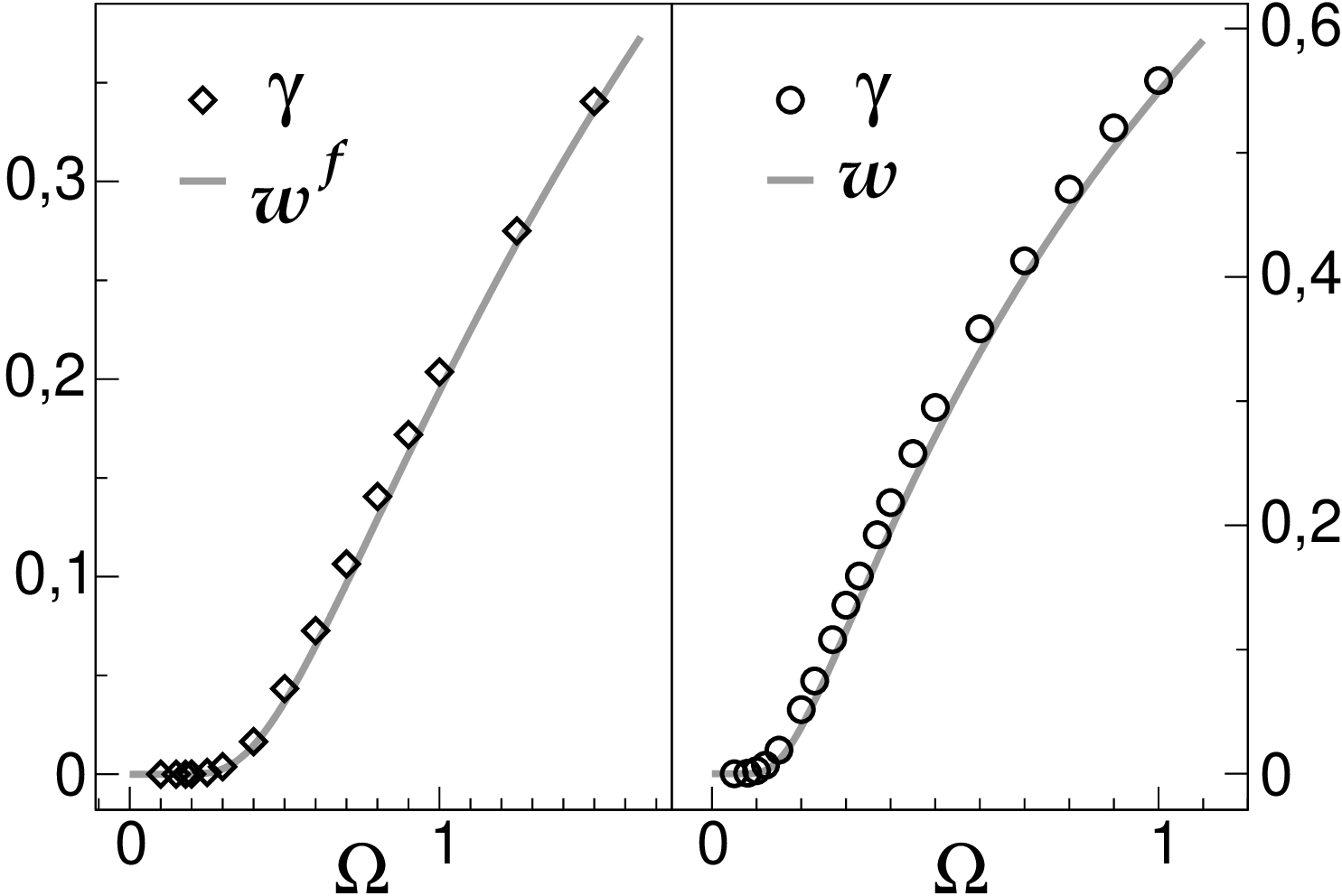}
\caption{\textit{Left plot}: the width of the first resonance (diamonds) compared with the pair production curve $w^f(\Omega)$ (solid line).\break
\textit{Right plot}: the width of the first resonance (circles) for a quadratic potential. The solid line, giving a very good fit,  is   $w(\Omega)=-\pi^{-1}\ln[1-\exp(-\pi/(4^2\Omega^{3/2}))]$, analogous to $w^f(\Omega)$. }
\label{prod}
\end{figure}

Let us now consider the specific cases of the potentials $U(x)={\mathcal{E}}\,|x|$ and $U(x)=(1/2)\,m\,\omega^2x^2$. For the linear potential we introduce
\begin{eqnarray}
y=(2m{\mathcal{E}})^{\frac 13}x\,,\quad\!\! \lambda= \Bigl(\frac{2m}{{\mathcal{E}}^2}\Bigr)^{\frac 13}\enr \,,\quad\!\!\Omega= \frac 1{4c^2}\Bigl(\frac{2\,\mathcal{E}}{m^2}\Bigr)^{\frac23}
\label{variabililineari}
\end{eqnarray}
and for the quadratic potential we let
\begin{eqnarray}
y=(m\omega)^\frac12\,x\,\quad \lambda=(2/\omega)\,\enr \,,\,\quad\Omega={\omega}/({4mc^2}),
\label{variabiliquadratiche}
\end{eqnarray}
so that in both cases the non-relativistic limit is obtained for $\Omega\rightarrow 0$. 
Equations (\ref{DiracEquation}),   (\ref{equaPhi}) and (\ref{equa2Phi1}) , then,  specify to
\begin{eqnarray}
&{}&\!\!\!\!\!\!\!\!\!\!\!\!\!\!\!\!\!\!\!\!\!\psi'(y)-\,{\Omega}^{\frac12}\bigl[\Lambda_n(y)\,i\sigma_y+(2\Omega)^{-1}\sigma_x\bigr]\psi(y)=0\label{equadim_psi} \spazio{0.6}\\
&{}&\!\!\!\!\!\!\!\!\!\!\!\!\!\!\!\!\!\!\!\!\!\phi'(y)-i\,{\Omega}^{\frac12}\bigl[\Lambda_n(y)\sigma_z\,+i(2\Omega)^{-1}\sigma_x\bigr]\phi(y)=0\label{equadim_phi} \spazio{0.6}\\
&{}&\!\!\!\!\!\!\!\!\!\!\!\!\!\!\!\!\!\!\!\!\!\phi''_1(y)+\bigl[\Omega\Lambda_n^2(y)+
 i\,\Omega^{\frac{1}{2}}ny^{n-1}-(4\Omega)^{-1}\bigr]\phi_1(y)=0
\label{equadim_phi1}
\end{eqnarray}
with $n=1,2$ and $\Lambda_n(y)=\lambda+1/(2\Omega)-y^n$.
When $n=1$ the normalizable solutions of (\ref{equadim_phi1}) with complex spectral parameter are known, \cite{Tit2}, and they are all proportional to the cylinder function $D_{i\tau}(-z)$, with $\tau=(2\Omega^{1/2})^{-3}$ and $z=(-4\Omega)^{1/4}\Lambda_1(y)$.
Carrying out the calculations previously described, it is straightforward to arrive to the expression for the density of the states  for equation (\ref{equadim_psi}). The general properties of $\rho(\lambda,\Omega)$, as in (\ref{rhodilambda}), can be appreciated by looking at the complete numerical results. First  we see the convergence to the Schr\"odinger levels when $\Omega\rightarrow 0$. For instance, the peaks of the first resonances for $\Omega=0.01$ are located at 1.0197, 2.3274, 3.2284, 4.0555, 4.7756: these should be compared with the first zeroes of the derivative of the Airy function and of the function itself (even and odd solutions respectively),  1.0190, 2.3384, 3.2482, 4.0884, 4.8201. Secondly,   in comparison with the non-relativistic case,  the spacing of corresponding resonances decreases for increasing values of the energy and of the interaction strength. In Fig.\ref{linearspectra} we present the plot of $\rho(\lambda,\Omega)$ for $n=1$, $\Omega=0.3$ and 1. The fit  of the density of the states by a sum   $\sum_{i=1}^Nc_i\,\lambda_{0i}^2\gamma_i((\lambda^2-\lambda_{0i}^2)^2+\lambda_{0i}^2\gamma_i^2)^{-1}$ of BW curves with the appropriate parameters  $\lambda_{0i}$, $\gamma_i$ and coefficients $c_i$, gives a perfect superposition with  $\rho(\lambda,\Omega)$.
In the left part of Fig.\ref{prod} we compare the width $\gamma$ of the first resonance with the pair production per unit length and time in a uniform electric field, \cite{Schw}, that  in  the variables (\ref{variabililineari}) reads 
\begin{equation}
w^f(\Omega)=-\pi^{-1}\ln\Bigl[1-\exp\Bigl(-\pi/(4\Omega^{3/2})\Bigr)\Bigr]
\label{produzionedicppie}
\end{equation}
As stated above, the excellent agreement, without any free parameter to be adjusted, proves that  pair production and line broadening are two different descriptions of a same physical situation. The minor differences, mainly for increasing $\Omega$, can be partly assigned to the description of the spectrum in terms of BW lines and partly to the fact that the Schwinger pair production is an effective one-loop calculation, possibly under-estimating the actual production rate \cite{Don}: one could make these small differences vanishing  not by exponential but only by power law corrections in $\Omega$.
\begin{figure}
\includegraphics*[height=4.8 cm,width=7.2cm]{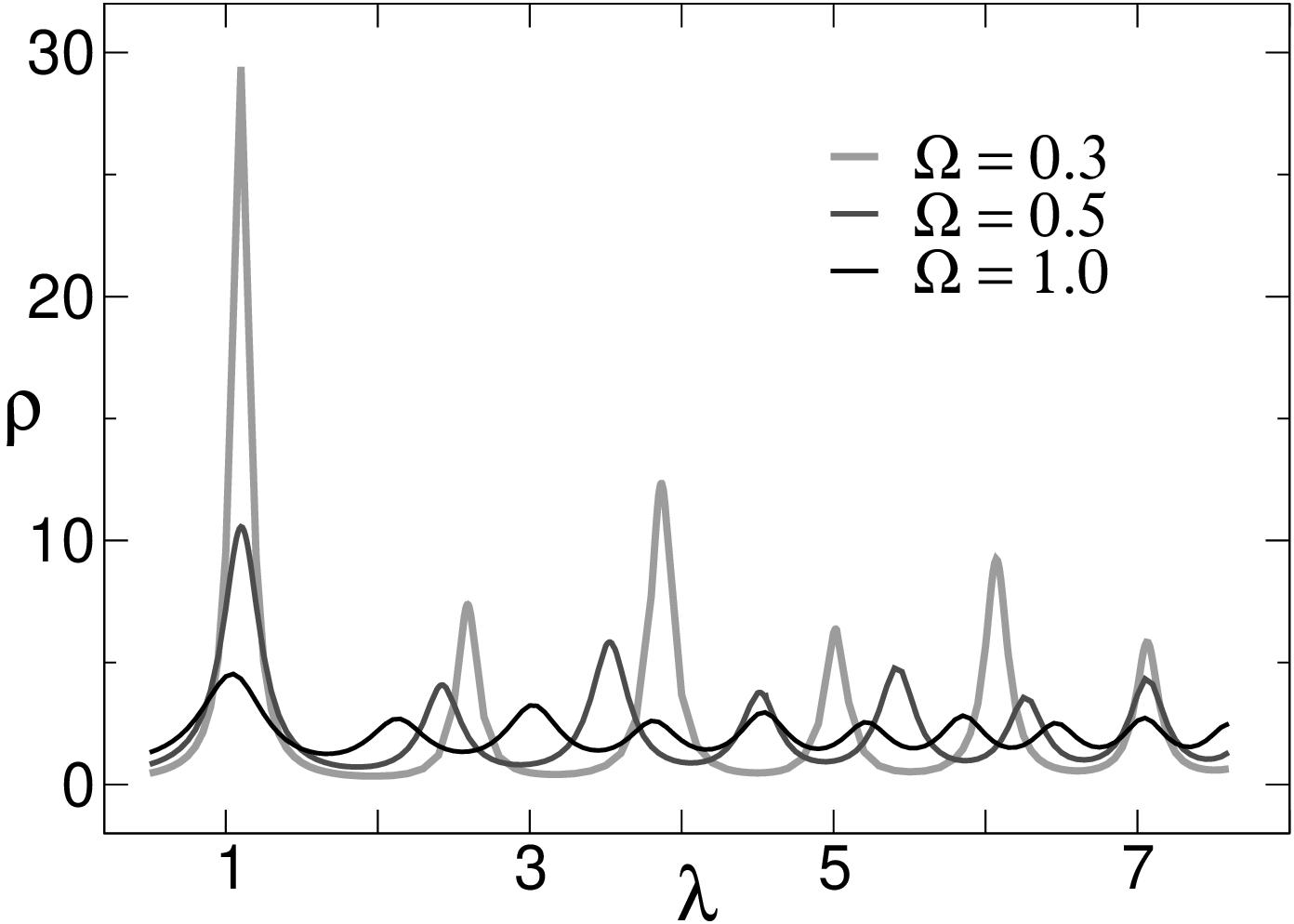}
\caption{$\rho(\lambda,\Omega)$ for the quadratic potential.}
\label{quadraticspectra}
\end{figure}

Similar considerations also apply to the quadratic potential
($n=2$ in (\ref{equadim_psi}-\ref{equadim_phi1})), although the computational technique is now different.
Actually, solutions of  (\ref{equadim_phi1}) exist in terms of  triconfluent Heun functions $H_T$, \cite{Duval},
$
\!\!\phi_1(y)\!=\!A\,e^{f}\,H_T(p,-3,q,z)\!+\!B\,e^{-f}\,H_T(p,3,q,-z)\,,
$
where  $f\!=\!-i\Omega^{\frac{1}{2}}\bigl(\lambda+(2\Omega)^{-1}-y^2/3\bigr)\,y$, $\,p\!=\!(3\,(4\Omega)^{-2})^\frac23$, $q\!=\!(12\,\Omega)^\frac13(\lambda+(2\Omega)^{-1})$ and $z\!=\!-i(2\,\Omega^\frac12/3)^\frac13\,y$.
Unfortunately no sufficient  informations on the asymptotic behavior of $H_T$ are available, to our knowledge, to determine the normalizable solutions for complex $\lambda$. Hence we use a completely numerical scheme that  extends to any potential $U(x)=a|x|^n$, for which analytical solutions do not  exist when $n\geq 3$.

\begin{figure}
\includegraphics*[height=4.8 cm,width=7.2cm]{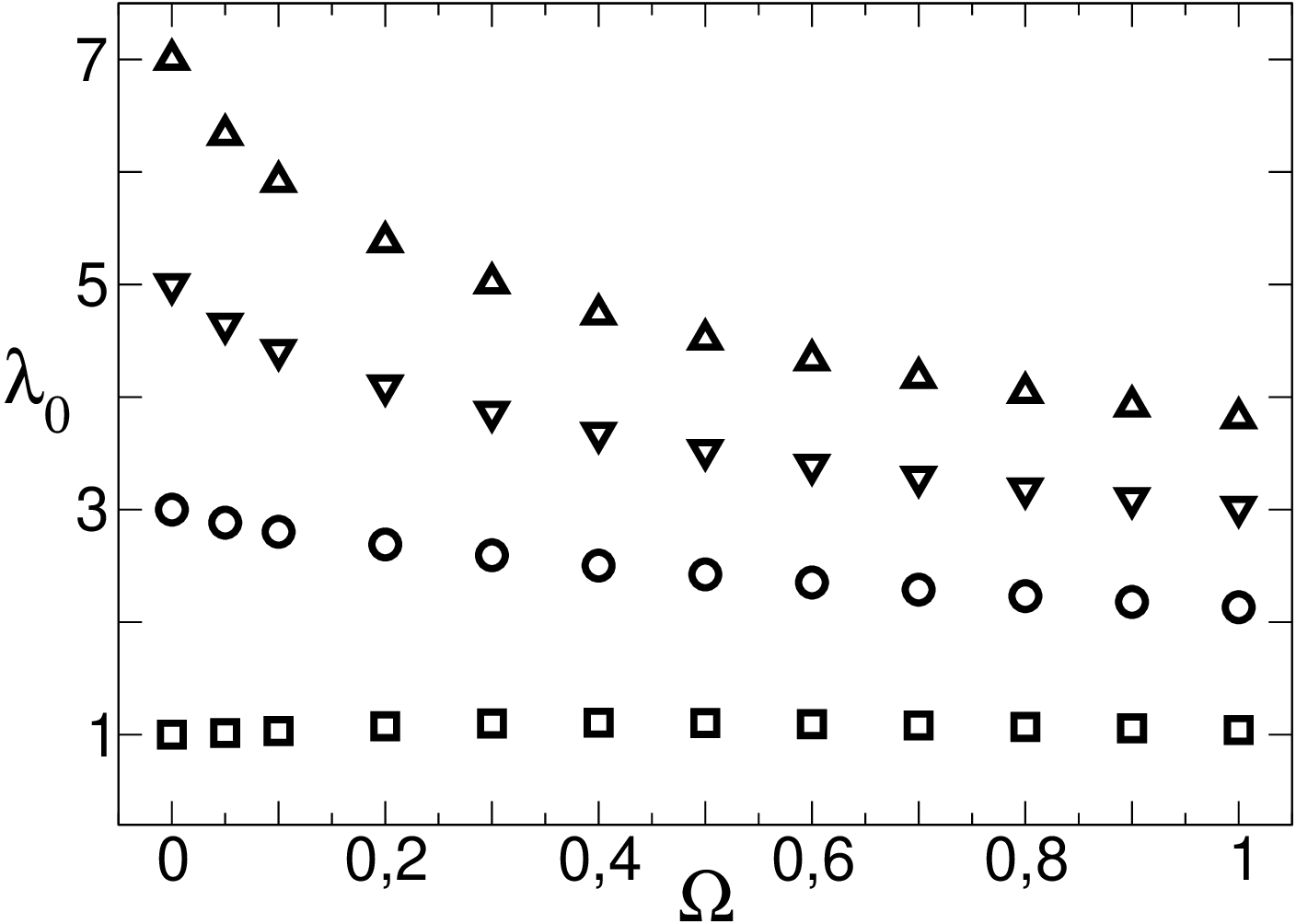}
\caption{The first BW maxima \textit{vs.} $\Omega$ for the quadratic potential. Non-relativistic bound states correspond to odd integers.}
\label{peaks}
\end{figure}

The calculations are straightforward and follow step by step the theory we have previously summarized. First we find, by numerical integration, a fundamental system of solutions of (\ref{equadim_phi}), from which we determine $\kappa(\lambda)$ according to (\ref{kappa}). Some care must be used in taking the limit, that is approached not in a monotonic but in an oscillating way, as it is evident by looking at the asymptotic leading terms of (\ref{equadim_phi1}): the convergence is increased by  constructing the sequence of the average points of pairs of nearby maxima and minima, whose limit is searched with  sufficiently high absolute and relative precision.  We then find the Weyl function $m(\lambda)$ from (\ref{rel_m_kappa}) and eventually, from (\ref{rhodilambda}), we deduce the density of the states, see Fig.\ref{quadraticspectra}.
The maxima of the first four BW lines are displayed in Fig.\ref{peaks}. Starting from the odd integers, that correspond to the non-relativistic values, we see that  their spacing decreases both for increasing $\Omega$ and $\lambda$: the same effect has been observed for the relativistic Landau levels, \cite{Kats}, and it loosely seems to repropose, in relativistic quantum mechanics,  the usual relationship between circular and harmonic motion. We can also remark that, as for the linear potential,  the lowest resonance has a central value that always remains near the non-relativistic value. The data of the higher resonances, instead,  are well fitted by decreasing exponentials in $\Omega$, they approach to each other and their unequal spacing should be taken into account in the construction of quarkonium models.
The right part of  Fig.\ref{prod} reproduces the width of the first BW resonance \textit{vs.} $\Omega$. According to what we said above, the
plot can give an estimate of the pair production for this case, not treated by QED. Remark that the data are well approximated by a curve $w(\Omega)$ very similar to the production in a constant field.

In conclusion, by using the possibilities offered by quantum mechanics, we have proved the continuity from the non-relativistic bound states to the states of the Dirac equation for an entire  class of  confining potentials, stressing  the fundamental role of the density of the states and clarifying  the  apparent physical contradictions of the absence of bound states for such potentials. Finally, although it is well known that the physical interpretation of the Dirac equation as a relativistic one particle wave equation presents many subtleties, we still believe that informations about its solutions and their  properties deserve a detailed knowledge.

%
%
%
%

\vfill\break

\end{document}